# PHYSICAL PICTURE OF THE INSURANCE MARKET

Amir Hossein Darooneh
Department of Physics, Zanjan University,
P. O. Box 45196-313, Zanjan, Iran.
darooneh@mail.znu.ac.ir

**Abstract:** We find the wealth distribution for an economic agent in the financial market, in analogy with standard derivation of generaliz Boltzman (Tsallis) factor in statistical mechanics. In this respect, Tsallis entropic index separates two different regimes, the large and small size market. The Pareto like wealth distribution is obtained in the case of small size market. A method for computing the premium is suggested based on the surplus average vanishing.

**Key Words:** Wealth distribution, Risk aversion, Tsallis Factor, Premium calculation principle.

## 1. Introduction:

Recently the physicists are interested to study the phenomena in other science disciplines by using the methods that are developed in physics. Among all branches of physics, the statistical mechanics appears as the most suitable context for exploring the dynamics of those phenomena that have stochastic nature. The economic problems have attracted the attention of physicists since last two decades [1, 2]. The price fluctuation in financial market is reason of this attraction. Predictions of prices and/or choosing the optimum strategies for trading are the most important problems in this respect.

The insurance is an important part of the financial market with respect to trading the risks. The study of insurance pricing strategies with the aid of ideas borrowed from statistical mechanics was begun by the work of author [3-8]. In this paper we extend the approach of Ref. 6 for finding the wealth distribution and computing the premium to small size markets. The next section is devoted to derive the wealth distribution which we call it the generalized Boltzman (Tsallis) factor. The effect of market size is discussed in terms of the wealth dependence of risk aversion parameter. In the section three a new method is suggested for premium calculation in agent-environment model of the insurance market [4-6]. In the last section we summarize all of the obtained results.

## 2. Generalized Boltzman (Tsallis) Factor

The financial market is combination of large number of economic agents which are interacting with each other through buying and/or selling. We consider the behavior of one of the agents for example an insurance company; all other agents may be regarded as its environment. The agent exchanges money when interacts with its environment. We suppose the financial market is a closed system and the clearing condition is satisfied. This means, the environment absorbs the money that the agent loses and will supply the agent's incomes.

$$W_a + W_e = W_0 = const. \quad (1)$$

The quantities $W_a$, $W_e$ and $W_0$ are the wealth of agent, its environment and total money in the market respectively.

There are many ways for the agent to possess $W_a$ amounts of money as a result of random loses and incomes in its trading. The quantity $\Gamma_a(W_a)$ represents number of these ways. This quantity may be corresponded with the concept of utility function in economic. The environment has also $\Gamma_e(W_e)$ ways to acquire $W_e$ amounts of money as its wealth. The state of the market is specified by two quantities $W_a$ and $W_e$, the market has $\Gamma_m(W_a, W_e)$ ways of reaching this specified state. Clearly,

$$\Gamma_m(W_a, W_e) = \Gamma_a(W_a)\Gamma_e(W_e) \quad (2)$$

By assumption we have, $\Gamma_X(0) = 1$, where subscript $X$ stands for the agent, environment or whole market.

Our common sense tells us, at any time the market chooses any one of these ways with equal probability because no reason exists for preferring some of them.

By definition, in equilibrium state the agent and its environment have most options for buying or selling. Existence of any restriction disturbs the equilibrium and decreases the number of ways that may be chosen for trading. Mathematically this means in the equilibrium state the function $\Gamma_m(W_a, W_e)$ should be maximized [9].

$$\frac{\partial \Gamma_m(W_a, W_e)}{\partial W_a} = 0 \quad (3)$$

The direct consequence of equilibrium condition is equality of risk aversion parameter for the agent and its environment [6].

$$\beta_a = \beta_e \quad (4)$$

Where the risk aversion parameter is defined as follows,

$$\beta_X = \frac{\partial \ln \Gamma_X(w)}{\partial w} \quad (5)$$

The probability for finding the agent in a specified state is [10],

$$p(W_a) = \frac{\Gamma_a(W_a)\Gamma_e(W_0 - W_a)}{\int_0^{W_0} \Gamma_a(w)\Gamma_e(W_0 - w)dw} \quad (6)$$

The wealth of environment is much larger than the agent's wealth hence the environment has much option for trading in comparison with the agent therefore we can write,

$$p(W_a) \propto \Gamma_e(W_0 - W_a) \quad (7)$$

The risk aversion parameter determines how an agent (environment) can trade in the market. The strategies for buying or selling are completely depends on this parameter. The environment risk aversion parameter may be function of its wealth.

$$\frac{d}{dW_e}(\frac{1}{\beta_e}) = Q \quad (8)$$

The left hand side of the above equation shows the reaction of environment to money exchange. In the first approximation we assume that it is a constant. For large environment any size of money exchange cannot affect the risk aversion parameter and $Q = 0$. Any variation in the wealth of small environment may cause the change in the risk aversion parameter; the case of non vanishing $Q$.

Here after we drop all indices correspond with environment for simplicity in notation.
The differential equation (8) has a simple solution,

$$\frac{1}{\beta} = \frac{1}{\beta_0} + QW \quad (9)$$

Where $\beta_0$ is constant of integration.
By substituting the above equation into equation (5) we obtain the following differential equation.

$$d \ln \Gamma(W) = \frac{\beta_0 dW}{1 + Q\beta_0 W} \quad (10)$$

The solution of this equation can be written in Q-deformed exponential function.

$$\Gamma(W) = \exp_Q(\beta_0 W) \quad (11)$$

Where,

$$\exp_Q(x) = (1 + Qx)^{1/Q} \quad (12)$$

In the limiting case, $Q \to 0$, it approaches to ordinary exponential form. Such function are appears in non extensive statistical mechanics that was introduced by Tsallis [11] and has many applications in all branches of science [12].
The wealth distribution may be written as,

$$\begin{aligned} p(W_a) &\propto \exp_Q[\beta_0(W_0 - W_a)] \\ &= \exp_Q(\beta_0 W_0) \exp_Q(-\beta W_a) \\ &\propto \exp_Q(-\beta W_a) \end{aligned} \quad (13)$$

Where,

$$\beta = \frac{\beta_0}{1 + Q\beta_0 W_0} \quad (14)$$

The wealth distribution (13) is called generalize Boltzman (Tsallis) factor and is familiar for all economists; this is a kind of Pareto (extreme value) distribution functions.

## 3. New Method for Premium Calculation

The insurance is a contract between insurer and insurant. Any happening loss incurred on the insurant party is covered by insurer, in return for money received as premium. In a competitive market the calculation of premium is so complicated. It is affected by random nature of risks and also by changing the number of insurants. Premium calculation principle is a rule that assign to any distribution function, corresponds with loss events, a real number [13]. It should also depend on market conditions [14] that are seen as variation in number of insurants [6].
Without lose of generality we can replace the wealth by the exchanged money in the equation 1, in this case the right hand side of the equation becomes zero due to clearing condition. The generalized Boltzman (Tsallis) factor is obtained again for the amount of money that an agent exchanges. In the insurance terminology, the sum of exchanged money for an insurer in the specified duration is called surplus. The surplus of an insurer may be written as,

$$S(t) = cI(t) - \sum_{j=1}^{N(t)} X_j \quad (15)$$

The insurer receives premium $c$ from $I(t)$ insurants and covers $N(t)$ loss results. In $j-th$ loss event, insurant charges the insurer for $X_j$ amount of money. All the variables $I(t)$, $N(t)$ and $X_j$'s are random. Their distribution functions are inferred from empirical data of the insurer.
The insurer naturally tries to increase its profit. When an insurance contract is over, its surplus should be positive or zero at least. This condition may be expressed as an average form.

$$\sum P(S(t))S(t) = 0 \quad (16)$$

The sum over all possible values for surplus is understood.
The $P(S(t))$ is escort probability and constructed from given distribution in equation (13).

$$P(S(t)) = \frac{p^{Q+1}(S(t))}{\sum p^{Q+1}(S(t))} \quad (17)$$

The escort probability preserves the additivity of the wealth averages and has also several advantages [15].
With respect to the above definition for escort probability the equation 16 becomes as,

$$\frac{\sum (cI(t)-Z(t))(1-Q\beta(cI(t)-Z(t)))^{\frac{Q+1}{Q}}}{\sum (1-Q\beta(cI(t)-Z(t)))^{\frac{Q+1}{Q}}} = 0 \quad (18)$$

This is a pure nonlinear equation and should be solved by numerical techniques or simulation. However it can be understood that in the limiting case (large market), $Q \to 0$, all the given results in Ref. 6 are obtained again. We are interested in realistic cases; the small market, that are not satisfied in prescribed condition and has non vanishing $Q$. From equation 18, it is understood that the premium depends on the number of insurants even it doesn't change in the time.

$$c = \frac{1}{I_0} \frac{\sum Z(t)(\exp_Q(\beta'(c)Z(t)))^Q}{\sum (\exp_Q(\beta'(c)Z(t)))^Q} \quad (19)$$

This is generalized form of the Esscher principle for computing the premium [13, 14]. The parameter $\beta'$ now is function of the premium amount.

$$\beta' = \frac{\beta}{1-Q\beta CI_0} \quad (20)$$

Where $I_0$ is the fixed number of insurants.

### 4. Conclusion

We retrieve the Pareto like wealth distribution for an agent by considering the agent–environment model for the financial market. This is what we called here as generalized Boltzman (Tsallis) factor. A new method for premium calculation is suggested on the basis of escort probability that is constructed from the above mentioned distribution. If we let the $Q$ approaches to zero, the previous results for large size market will be retrieved. In the small market the number of insurants is important even it is fixed in the time.